\documentclass{article}
\usepackage[utf8]{inputenc}
\usepackage{fullpage}
\usepackage{float}
\usepackage{amsmath,amsthm, amsfonts}
\usepackage{graphicx,caption}
\usepackage{xcolor}
\usepackage{authblk}
\usepackage{arydshln}
\usepackage{pifont}
\usepackage{comment}

\newcommand{\pkg}[1]{{\fontseries{b}\selectfont #1}}

\newcommand{\keywords}[1]{\textbf{\textit{Keywords---}} #1}

\usepackage[citestyle = numeric, bibstyle =  apa, sorting = none]{biblatex}
\makeatletter
\RequireBibliographyStyle{numeric}

\title{metamedian: An R package for meta-analyzing studies reporting medians}
\author[1]{Sean McGrath}
\author[2]{XiaoFei Zhao}
\author[3]{Omer Ozturk}
\author[4]{Stephan Katzenschlager}
\author[5]{Russell Steele}
\author[6,7,8]{Andrea Benedetti}
\date{}

\affil[1]{\small Department of Biostatistics, Harvard T.H. Chan School of Public Health, Boston, MA, USA}
\affil[2]{\small MOE Key Laboratory of Bioinformatics, Bioinformatics Division, BNRIST and Department of Automation, Tsinghua University, Beijing, China}
\affil[3]{\small Department of Statistics, The Ohio State University, Columbus, OH, USA}
\affil[4]{\small Department of Anesthesiology, Heidelberg University Hospital, Heidelberg, Germany}
\affil[5]{\small Department of Mathematics and Statistics, McGill University, Montreal, Quebec, Canada}
\affil[6]{\small Department of Epidemiology, Biostatistics, and Occupational Health, McGill University, Montreal, Quebec, Canada}
\affil[7]{\small Respiratory Epidemiology and Clinical Research Unit (RECRU), McGill University Health Centre, Montreal, Quebec, Canada}
\affil[8]{\small Department of Medicine, McGill University, Montreal, Quebec, Canada}

\begin{document}

\maketitle

\begin{abstract}
When performing an aggregate data meta-analysis of a continuous outcome, researchers often come across primary studies that report the sample median of the outcome. However, standard meta-analytic methods typically cannot be directly applied in this setting. In recent years, there has been substantial development in statistical methods to incorporate primary studies reporting sample medians in meta-analysis, yet there are currently no comprehensive software tools implementing these methods. In this paper, we present the \pkg{metamedian} R package, a freely available and open-source software tool for meta-analyzing primary studies that report sample medians. We summarize the main features of the software and illustrate its application through real data examples involving risk factors for a severe course of COVID-19. 
\end{abstract}


\section{Introduction} \label{sec: intro}

An aggregate data meta-analysis combines summary statistics of an outcome of interest from multiple studies. When the outcome of interest is continuous, statistical methods for performing an aggregate data meta-analysis often assume that the primary studies report the sample mean and standard deviation of the outcome. However, primary studies may report the sample median of the outcome rather than the sample mean, which commonly occurs when the distribution of the outcome is skewed \cite{higgins2020cochrane}. 

Standard meta-analytic methods typically cannot be directly applied when primary studies report sample medians. Consequently, a number of statistical methods have been developed to incorporate studies reporting sample medians in meta-analysis. These methods can be classified into two groups: mean-based methods and median-based methods. Mean-based methods impute the sample means and standard deviations of the outcome from primary studies that report sample medians in order to apply standard meta-analytic methods based on the (imputed) study-specific sample means and standard deviations. Many authors have proposed sample mean and standard deviation estimators for this context \cite{hozo2005estimating,bland2015estimating, wan2014estimating, kwon2015simulation, luo2018optimally, mcgrath2020estimating, shi2020optimally, shi2020estimating, rychtavr2020estimating, walter2022estimation, cai2021estimating, yang2021generalized, balakrishnan2022unified}. Median-based methods directly meta-analyze the study-specific sample medians in order to estimate a pooled median of the outcome or the difference of medians across groups \cite{mcgrath2019one, mcgrath2020meta, ozturk2020meta}. 

A software tool implementing these methods can facilitate performing high-quality meta-analyses for a few reasons. First, some of these methods can be challenging and laborious to apply without available software (e.g., \cite{mcgrath2020estimating, cai2021estimating, mcgrath2020meta} require numerically solving ad-hoc optimization problems), which may limit their adoption in practice. Second, it can be insightful to perform sensitivity analyses where data analysts apply several of these methods and evaluate how the conclusions of their meta-analysis change, as these methods are based on different assumptions and estimation strategies. A comprehensive software tool implementing these methods in a standardized way can facilitate performing such sensitivity analyses. 

In this paper, we present the \pkg{metamedian} R package \cite{metamedian}, a freely available and open-source software tool for meta-analyzing primary studies that report sample medians. The \pkg{metamedian} R package is available on the Comprehensive R Archive Network (CRAN) at \url{https://CRAN.R-project.org/package=metamedian}, and the development version of the package is available on GitHub at \url{https://github.com/stmcg/metamedian}. The package implements both mean-based methods and median-based methods. When applicable, the widely used \pkg{metafor} package \cite{metafor} is internally applied when pooling the study-specific estimates, which gives users a wide array of pooling options (e.g., supporting common effect and random effects models, various between-study heterogeneity estimators, meta-regression analyses, etc.) and facilitates performing a number of subsequent analyses (e.g., generating forest plots and funnel plots, testing small study effects, etc.). 

The remainder of the paper is structured as follows. We briefly summarize the literature on statistical methods for meta-analyzing studies reporting sample medians. We then describe the main features of the software and illustrate its application in a real-life meta-analysis. We conclude with a discussion of related software, limitations, and future directions in the development of \pkg{metamedian}.

\section{Methods} \label{sec: methods}

\subsection{Standard meta-analytic methods} \label{sec: standard methods}

We begin by recalling some standard statistical models and estimators for an aggregate data meta-analysis of a continuous outcome. These models and estimators are the foundation of many of the methods implemented in the \pkg{metamedian} package. 

Suppose that the meta-analysis consists of $K$ primary studies. Let $y_k$ denote the estimate of the outcome measure in the $k\textsuperscript{th}$ primary study (e.g., the difference of sample means across two arms in a trial), and let $\theta_k$ denote the true outcome measure in the $k\textsuperscript{th}$ primary study. It is assumed that
\begin{equation*}
    y_k \sim \text{Normal}(\theta_k, \sigma^2_k),
\end{equation*}
where the within-study sampling variances, $\sigma^2_k$, are considered to be known. The common effect model assumes that the true outcome measures of the primary studies are equal  (i.e., $\theta_k = \theta$ for all $k$). The random effects model assumes that the true outcome measures of the primary studies differ and are distributed as
\begin{equation*}
    \theta_k \sim \text{Normal}(\theta, \tau^2).
\end{equation*}
Throughout, we refer to $\theta$ as the pooled outcome measure and $\tau^2$ as the between-study variance. 

The classic inverse-variance weighted estimator of the pooled outcome measure and an estimate of its standard error (SE) are given by
\begin{equation*}
    \hat{\theta} = \frac{\sum_{k = 1}^K w_k y_k}{\sum_{k = 1}^K w_k}, \qquad \widehat{\text{SE}}(\hat{\theta}) = \sqrt{\frac{1}{\sum_{k = 1}^K w_k}}, 
\end{equation*}
where the weights, $w_k$, depend on the type of meta-analytic model used. In a common effect meta-analysis, $w_k = \frac{1}{\sigma_k^2}$. In a random effects meta-analysis, $w_k = \frac{1}{\sigma_k^2 + \hat{\tau}^2}$ where $\hat{\tau}^2$ denotes an estimate of the between-study variance. See \cite{viechtbauer2005bias, veroniki2016methods, langan2019comparison} for discussions of estimators of the between-study variance.

\subsection{Methods to meta-analyze studies reporting medians}

There are two key challenges to applying standard meta-analytic methods when primary studies report the sample median of the outcome. One challenge is that studies reporting the sample median of the outcome often do not report an estimate of its SE (i.e., $\sigma_k$), which is needed to compute the weights in the inverse-variance weighted estimator of the pooled outcome measure. Instead, studies reporting sample medians commonly report the first and third quartiles and/or minimum and maximum values of the outcome. In other cases, these studies may not report any other summary statistics of the outcome besides the sample median, which typically occurs when the outcome of interest (in the meta-analysis) is not one of the primary outcomes in the study. 

A second challenge is that some primary studies may report estimates of the mean of the outcome whereas other primary studies may report estimates of the median of the outcome. In general, outcome measures based on means (e.g., the difference of means across groups) are not equal to outcome measures based on medians (e.g., the difference of medians across groups). To apply standard meta-analytic methods, all primary studies must contribute an estimate of the same outcome measure.

The following subsections describe statistical methods developed to address these challenges. 

\subsubsection{Mean-based methods}

The most commonly applied approach to meta-analyze studies reporting sample medians involves imputing sample means and standard deviations of the outcome from studies reporting medians \cite{hozo2005estimating,bland2015estimating, wan2014estimating, kwon2015simulation, luo2018optimally, mcgrath2020estimating, shi2020optimally, shi2020estimating, rychtavr2020estimating, walter2022estimation, cai2021estimating, yang2021generalized, balakrishnan2022unified}. Then, data analysts may apply standard meta-analytic methods based on the (imputed) study-specific sample means and standard deviations. For instance, this approach may be used to impute the difference of sample means and its usual SE estimate from studies reporting sample medians in order to meta-analyze the difference of means. Because these methods estimate an outcome measure based on means, we refer to these methods as mean-based methods. 

The literature on mean-based methods is growing large, which we very briefly summarize next. These methods typically consider that a primary study may report some of the following statistics of the outcome: minimum value ($q_{\text{min}}$), first quartile ($q_1$), median ($q_2$), third quartile ($q_{3}$), maximum value ($q_{\text{max}}$), and sample size ($n$). In particular, they typically consider that one of the following sets of summary statistics is reported:
\begin{align*}
    S_1 & = \{q_{\text{min}}, q_{2}, q_{\text{max}}, n \} \\
    S_2 & = \{q_{1}, q_{2}, q_{3}, n \} \\
    S_3 & = \{q_{\text{min}}, q_{1}, q_{2}, q_{3}, q_{\text{max}}, n \}. 
\end{align*}
Hozo et al.\ \cite{hozo2005estimating} were amongst the first to propose and systematically study mean-based methods, whereby they developed sample mean and standard deviation estimators from $S_1$. After, Bland \cite{bland2015estimating} proposed corresponding estimators from $S_3$, and Wan et al.\ \cite{wan2014estimating} proposed estimators from $S_2$ along with other estimators from $S_1$ and $S_3$ under the assumption that the outcome is normally distributed. Since then, Luo et al.\ \cite{luo2018optimally} and Shi et al.\ \cite{shi2020optimally} developed new estimators to achieve certain optimality properties under the assumption that the outcome is normally distributed. Motivated by the observation that studies often report sample medians instead of sample means because the distribution of the outcome is skewed, Kwon and Reis \cite{kwon2015simulation}, McGrath et al. \cite{mcgrath2020estimating}, Shi et al. \cite{shi2020estimating}, and Cai et al. \cite{cai2021estimating} developed estimators for skewed data. Other estimators have been developed which require the data analysts to specify the assumed parametric distribution of the outcome, such as the normal or log-normal distribution \cite{yang2021generalized, balakrishnan2022unified}. 

While most of the literature on mean-based methods has focused on better estimating the sample mean and standard deviation from $S_1$, $S_2$, or $S_3$, another important consideration is the performance of such methods in the context of meta-analysis. McGrath et al.\ \cite{mcgrath2023standard} showed that using the imputed study-specific sample means and standard deviations in place of the actual (unreported) sample means and standard deviations in standard meta-analytic methods can severely underestimate the within-study SEs for studies reporting medians, which can result in negative downstream consequences in meta-analysis. Moreover, they described a bootstrap approach to better estimate the within-study SEs when using the mean estimators of McGrath et al.\ \cite{mcgrath2020estimating} and Cai et al.\ \cite{cai2021estimating}. Yang et al.\ \cite{yang2021generalized} and Balakrishnan et al.\ \cite{balakrishnan2022unified} also provided estimators of the within-study SE that appropriately account for the variability of their mean estimators.

\subsubsection{Median-based methods}

Another line of the literature has focused on developing methods to directly meta-analyze the study-specific sample medians, i.e., without first imputing sample means and standard deviations \cite{mcgrath2019one, mcgrath2020meta, ozturk2020meta}. These methods estimate the pooled median of the outcome when the meta-analysis consists of one-group studies, and they estimate the pooled difference of medians in the case of two-group studies (e.g., studies with treatment and control groups). We refer to these methods as median-based methods because they estimate an outcome measure based on medians.

The main distinction between the different median-based methods is how they address the challenge of unreported within-study SEs. McGrath et al.\ \cite{mcgrath2019one, mcgrath2020meta} considered methods that avoid the need for within-study SEs altogether. Specifically, they considered an approach that takes the median of the study-specific estimates as the point estimate of the pooled outcome measure. For instance, in a meta-analysis of one-group studies, this approach uses the median of the study-specific medians as the pooled estimate. For two-group studies, this approach uses the median of the study-specific differences of medians as the pooled estimate. A nonparametric confidence interval around the pooled estimate is obtained by taking suitable quantiles of the study-specific estimates. Weighted versions (based on the study-specific sample sizes) were also considered. 

McGrath et al.\ \cite{mcgrath2020meta} and Ozturk and Balakrishnan \cite{ozturk2020meta} considered methods that estimate the within-study SEs in order to perform an inverse-variance weighted meta-analysis. Specifically, McGrath et al.\ \cite{mcgrath2020meta} proposed parametric estimators of the within-study SE from $S_1$, $S_2$, or $S_3$ summary statistics based on an estimation strategy referred to as Quantile Matching Estimation (QE). To distinguish this median-based method using QE from the mean-based method using QE \cite{mcgrath2020estimating} (see Section \ref{sec: software mean-based methods}), we refer to the median-based method as $\text{QE}_{\text{median}}$ and the mean-based method as $\text{QE}_{\text{mean}}$. Ozturk and Balakrishnan \cite{ozturk2020meta} proposed nonparametric estimators of the within-study SE from $S_2$ summary statistics as well as other sets of summary statistics (see Appendix A for details), which they referred to as the Confidence Distribution (CD) approach.

When the meta-analysis consists of some primary studies that report the sample median of the outcome and other primary studies that report the sample mean, all of these methods assume that the distribution of the outcome is symmetric in the primary studies that report sample means. That is, they assume that the mean of the outcome equals the median in such primary studies.

\subsubsection{Comparison of methods} \label{sec: comparison}

Given the number of methods developed for meta-analyzing studies reporting medians, data analysts may ask how to choose the most suitable methods for their applications. In this subsection, we summarize some comparisons of these methods performed in previous studies. 

First, consider the comparison between mean-based methods and median-based methods. Since mean-based methods and median-based methods estimate different outcome measures, one consideration is which outcome measure is most appropriate for the application at hand. For instance, if the distribution of the outcome is highly skewed, one may find that the mean of the outcome distribution is not a very meaningful outcome measure for the application and instead prefer to estimate the median. A more detailed discussion comparing mean-based outcome measures and median-based outcome measures in this context is given in McGrath et al.\ \cite{mcgrath2020meta}. Apart from considerations on the outcome measure, two factors that strongly differentiate the performance of mean-based methods and median-based methods are (i) the proportion of primary studies reporting medians and (ii) the skewness of the outcome distribution in the primary studies. For (i), as the proportion of primary studies reporting medians increases, the performance of median-based methods often improves and the performance of mean-based methods often worsens \cite{mcgrath2019one, mcgrath2020meta, mcgrath2023standard}, as one may expect. For (ii), when the outcome distribution is highly skewed, mean-based methods can perform poorly \cite{mcgrath2019one, mcgrath2020meta,mcgrath2023standard}. In practice, data analysts can use Bowley's coefficient of skewness \cite{bowley1901} to evaluate the skewness of the outcome in a primary study, as it only depends on $S_2$ summary statistics (e.g., see \cite{mcgrath2019one, mcgrath2020meta, ozturk2020meta, mcgrath2023standard}).

Second, consider the comparison of the various mean-based methods. Most simulation studies evaluating the performance of mean-based methods found that the performance of the mean and standard deviation estimators strongly depends on the underlying distribution of the outcome. If the outcome is normally distributed in the primary studies, then the methods that assume normality (e.g., Wan et al.\ \cite{wan2014estimating}, Luo et al.\ \cite{luo2018optimally}, Shi et al.\ \cite{shi2020optimally}, Yang et al.\ \cite{yang2021generalized}) often perform best. However, when the distribution of the outcome is not normal (especially skewed), the methods developed under more flexible distributional assumptions (e.g., McGrath et al.\ \cite{mcgrath2020estimating}, Cai et al.\ \cite{cai2021estimating}) often perform best. Additionally, regardless of the approach used to estimate the study-specific means, using a within-study SE estimator that appropriately accounts for the variability of the mean estimators (i.e., the parametric bootstrap \cite{mcgrath2023standard} and plug-in approaches \cite{yang2021generalized}) can yield better inference at the meta-analysis level compared to the naïve SE estimator \cite{mcgrath2023standard}.

Last, consider the comparison of the median-based methods. An advantage of the approach based on taking the median of the study-specific outcome measure estimates \cite{mcgrath2019one, mcgrath2020meta} is that it is able to incorporate studies that only report the sample median of the outcome (rather than $S_1$, $S_2$, $S_3$, or other sets of summary statistics). However, since the $\text{QE}_{\text{median}}$ \cite{mcgrath2020meta} and CD \cite{ozturk2020meta} methods meta-analyze studies using inverse-variance weighting, advantages of these approaches include the following: they are able to estimate between-study heterogeneity; they are typically more efficient; and they allow data analysts to perform a number of standard subsequent analyses (e.g., generating forest plots and funnel plots, testing small study effects).

\section{Software functionality} \label{sec: software}

To demonstrate the software's functionality, we will use as an example a meta-analysis recently performed by Katzenschlager et al.\ \cite{katzenschlager2021can} that aimed to identify risk factors for a severe course of COVID-19. The analysis we focus on compared the average age between COVID-19 infected patients who died and those who survived. All 52 primary studies either reported $S_2$ summary statistics of age in both groups of patients or reported the mean, standard deviation, and sample size in both groups. The \pkg{metamedian} package includes two data sets for this example: \verb|dat.age_raw| contains the extracted summary data for all 52 primary studies, and \verb|dat.age| contains the summary study for all primary studies except one which had a very small sample size in the nonsurvivor group (see Section \ref{sec: examples} for further details).

We chose this example because all of the methods in the \pkg{metamedian} package give similar results and lead to the same conclusions. We believe that using an example where different methods lead to different conclusions could potentially distract readers from the primary focus of the paper (i.e., describing the utility of the \pkg{metamedian} package). However, for interested readers, we provide analyses of two other outcome variables (i.e., aspartate transaminase and creatine kinase levels) from the meta-analysis of Katzenschlager et al.\ \cite{katzenschlager2021can} in Appendix C. In these additional examples, some methods lead to noticeably different results. 

The \pkg{metamedian}  package can be installed from CRAN and loaded by running the following commands in the R console: 
\begin{verbatim}
install.packages("metamedian")
library("metamedian")
\end{verbatim}

The following subsections describe the requirements on the input data set and the key functions in the \pkg{metamedian} package.

\subsection{Data set} \label{sec: data set}

The key functions in the \pkg{metamedian} package require users to supply an input data set, \verb|data|, containing the summary data extracted from the primary studies. The input data set must be a data frame, where each row corresponds to a primary study in the meta-analysis. For one-group studies, the input data set can include the following columns: \verb|min.g1| (minimum value), \verb|q1.g1| (first quartile), \verb|med.g1| (median), \verb|q3.g1| (third quartile), \verb|max.g1| (maximum value), \verb|n.g1| (sample size), \verb|mean.g1| (mean), and \verb|sd.g1| (standard deviation). For two-group studies, the input data set can additionally include the following columns for the summary data in the second group: \verb|min.g2|, \verb|q1.g2|, \verb|med.g2|, \verb|q3.g2|, \verb|max.g2|, \verb|n.g2|, \verb|mean.g2|, and \verb|sd.g2|. When constructing the input data set for two-group studies, note that the outcome measure (i.e., difference of means or the difference of medians across groups) is based on the group 1 value minus the group 2 value. If the study does not report one of the summary statistics, the relevant entry in the input data set must be set to \verb|NA|. Some additional columns can be included in the input data set when using some methods (e.g., the CD method \cite{balakrishnan2022unified}), which are detailed in Appendix A.

As an example, the data set \verb|dat.age| corresponding to our example is formatted in this manner. The first five rows of \verb|dat.age| (excluding the \verb|author| column corresponding to the authors of the primary studies) are displayed below. The group 1 values correspond to the ages (in years) in the nonsurvivor group, and the group 2 values correspond to the ages in the survivor group.
\begin{verbatim}
   n.g1 q1.g1 med.g1 q3.g1 mean.g1 sd.g1 n.g2 q1.g2 med.g2 q3.g2 mean.g2 sd.g2
1    35    NA     NA    NA   77.00  8.30  157    NA     NA    NA   65.60 15.60
2    32    NA     NA    NA   64.60 11.20   20    NA     NA    NA   51.90 12.90
3    31  65.0   76.0 82.00      NA    NA  122  56.0   63.0  69.0      NA    NA
4    16  68.0   72.0 81.00      NA    NA  137  56.0   63.0  70.0      NA    NA
5    29    NA     NA    NA   63.60 17.68  351    NA     NA    NA   52.85 16.35
\end{verbatim}

\subsection{Main functions}

\subsubsection{Mean-based methods} \label{sec: software mean-based methods}

The \verb|metamean| function performs a meta-analysis using mean-based methods. For one-group studies, this function meta-analyzes the mean of the outcome. For two-group studies, this function meta-analyzes the difference of means across the two groups.

The method for estimating the mean(s) of the outcome in a primary study reporting $S_1$, $S_2$, or $S_3$ summary statistics is specified by the \verb|mean_method| argument. The argument \verb|mean_method| can either be a vector, in which case the $k\textsuperscript{th}$ element is the method used for the $k\textsuperscript{th}$ primary study in the input data set, or a scalar, in which case the same method is used for all primary studies. The options for \verb|mean_method| are listed below:
\begin{itemize}
    \item \verb|"wan"|: This option applies the mean estimator of Hozo et al.\ \cite{hozo2005estimating} in scenario $S_1$, Wan et al.\ \cite{wan2014estimating} in scenario $S_2$, and Bland \cite{bland2015estimating} in scenario $S_3$.
    \item \verb|"luo"|: This option applies the mean estimator of Luo et al.\ \cite{luo2018optimally} in all three scenarios.
    \item \verb|"shi_normal"|: This option applies the mean estimator of Luo et al.\ \cite{luo2018optimally} in scenarios $S_1$ and $S_2$ and the mean estimator of Shi et al. \cite{shi2020optimally} in scenario $S_3$.
    \item \verb|"shi_lognormal"|: This option applies the mean estimator of Shi et al.\ \cite{shi2020estimating} in all three scenarios.
    \item \verb|"qe"|: This option applies the $\text{QE}_{\text{mean}}$ mean estimator of McGrath et al.\ \cite{mcgrath2020estimating} in all three scenarios.
    \item \verb|"bc"|: This option applies the Box-Cox (BC) mean estimator of McGrath et al.\ \cite{mcgrath2020estimating} in all three scenarios.
    \item \verb|"mln"|: This option applies the Method for Unknown Non-Normal Distributions (MLN) mean estimator of Cai et al.\ \cite{cai2021estimating} in all three scenarios.
    \item \verb|"yang"|: This option applies the mean estimator of Yang et al.\ \cite{yang2021generalized} under the assumption of normality in all three scenarios.
\end{itemize}
Note that some of these options involve different mean estimators in different scenarios. This is because those papers did not propose mean estimators in all three scenarios (e.g., Wan et al.\ \cite{wan2014estimating}), and instead recommended existing approaches in the scenarios they did not consider. 

The method for estimating the SE of the mean estimators in scenarios $S_1$, $S_2$, and $S_3$ is specified by the \verb|se_method| argument. The options include the following:
\begin{itemize}
    \item \verb|"naive"|: This option applies the naïve SE estimator, which uses the estimated standard deviation divided by the square root of the sample size. This option is available when using any of the mean estimators. The approach used to estimate the standard deviation is specified by the \verb|sd_method| argument, which can be set to \verb|"wan"| (Wan et al.\ \cite{wan2014estimating}), \verb|"shi_normal"| (Shi et al.\ \cite{shi2020optimally}), \verb|"shi_lognormal"| (Shi et al.\ \cite{shi2020estimating}), \verb|"qe"| (McGrath et al.\ \cite{mcgrath2020estimating}, $\text{QE}_{\text{mean}}$ approach), \verb|"bc"| (McGrath et al.\ \cite{mcgrath2020estimating}, BC approach), \verb|"mln"| (Cai et al.\ \cite{cai2021estimating}, MLN approach), and \verb|"yang"| (Yang et al.\ \cite{yang2021generalized}). Note that the \verb|"shi_normal"| option uses the standard deviation estimators of Wan et al.\ \cite{wan2014estimating} in scenarios $S_1$ and $S_2$. 
    \item \verb|"bootstrap"|: This option applies the parametric bootstrap approach described by McGrath et al.\ \cite{mcgrath2023standard}. This option is available when using the $\text{QE}_{\text{mean}}$, BC, and MLN mean estimators. The number of bootstrap replicates is specified by the \verb|nboot| argument, which is set to \verb|1000| by default. Since the bootstrap approach involves random sampling, users may wish to set a random number seed to ensure reproducibility (e.g., running \verb|set.seed(1234)| in the R console) when using this option.
    \item \verb|"plugin"|: This option uses the analytically derived SE of the mean estimator, and plugs in the estimated standard deviation in place of the distributional standard deviation. This option is available when using the Yang et al.\ \cite{yang2021generalized} mean estimator.
\end{itemize}

When a primary study reports the sample mean and standard deviation of the outcome, the sample mean is used as the point estimate and the standard deviation divided by the square root of the sample size is used as the SE estimate. For two-group studies, the outcome measure estimate is the difference in sample means and its SE estimate is the square root of the of the sum of the squared SE estimates in both groups. 

After obtaining study-specific outcome measure estimates and their SE estimates, an inverse-variance weighted meta-analysis is performed via the \verb|rma.uni| function in the \pkg{metafor} package. By default, a random effects meta-analysis is performed where the Restricted Maximum Likelihood (REML) approach is used to estimate between-study heterogeneity. Additional arguments supplied to the \verb|metamean| function are directly passed to the \verb|rma.uni| function so that users can specify how to pool the study-specific estimates, such as specifying whether a common effect or random effects model is used or including effect modifier variables (i.e., performing a meta-regression). The output of the \verb|rma.uni| function is returned, giving users the option of performing subsequent analyses using functions available in the \pkg{metafor} package.

As an example, consider the following application of the \verb|metamean| function to the data set from our example (i.e., \verb|dat.age|). We can perform a random effects meta-analysis where we apply the MLN method to estimate the study-specific means and the parametric bootstrap approach (with 100 bootstrap replicates for ease of computation) to estimate their SEs as follows:
\begin{verbatim}
set.seed(1234)
metamean(data = dat.age, mean_method = "mln", se_method = "bootstrap", nboot = 100)
\end{verbatim}
The output is given below.
\begin{verbatim}
Random-Effects Model (k = 51; tau^2 estimator: REML)

tau^2 (estimated amount of total heterogeneity): 28.2204 (SE = 7.0657)
tau (square root of estimated tau^2 value):      5.3123
I^2 (total heterogeneity / total variability):   86.75%
H^2 (total variability / sampling variability):  7.55

Test for Heterogeneity:
Q(df = 50) = 362.9420, p-val < .0001

Model Results:

estimate      se     zval    pval    ci.lb    ci.ub      
 12.8410  0.8367  15.3464  <.0001  11.2010  14.4809  *** 

---
Signif. codes:  0 ‘***’ 0.001 ‘**’ 0.01 ‘*’ 0.05 ‘.’ 0.1 ‘ ’ 1
\end{verbatim}
The output indicates that the pooled estimate of the difference of means (mean age in the nonsurvivor group minus the mean age in the survivor group) is 12.84 years [95\% CI: 11.20, 14.48]. 

\subsubsection{Median-based methods}

The \verb|metamedian| function performs a meta-analysis using median-based methods. These methods meta-analyze the median of the outcome for one-group studies and the difference of medians for two-group studies. 

The argument \verb|method_median| is a scalar that specifies the method used to perform the meta-analysis. The options are listed below:
\begin{itemize}
    \item \verb|"mm"|: This option applies the Median of Medians (MM) method \cite{mcgrath2019one} when the meta-analysis consists of one-group primary studies and applies the Median of the Difference of Medians (MDM) method \cite{mcgrath2020meta} when the meta-analysis consists of two-group primary studies. These methods only require that \verb|data| contain the median or mean of the outcome (in both groups) in the primary studies. 
    \item \verb|"wm"|: This option applies weighted versions of the methods described for the \verb|"mm"| option, where studies are weighted proportional to their total sample size \cite{mcgrath2019one, mcgrath2020meta}. 
    \item \verb|"qe"|: This option applies the $\text{QE}_{\text{median}}$ method \cite{mcgrath2020estimating}, which is applicable whether the meta-analysis consists of one-group studies or two-group studies. Recall that this method parametrically estimates the SEs of the study-specific medians (for one-group studies) or differences of medians (for two-group studies). Then, an inverse-variance weighted meta-analysis is performed via the \verb|rma.uni| function in the \pkg{metafor} package in the same manner as described for the mean-based methods. The output of the \verb|rma.uni| function is returned. This method requires that \verb|data| contains the study-specific $S_1$, $S_2$, $S_3$ summary statistics of the outcome or the sample mean, standard deviation, and sample size (in both groups). 
    \item \verb|"cd"|: This option applies the CD method \cite{ozturk2020meta}, which is applicable for one-group studies. Recall that this method nonparametrically estimates the SEs of the study-specific medians. Then, an inverse-variance weighted meta-analysis is performed where a Jackknife approach is used to estimate the variance of the pooled outcome measure estimate. This method requires that \verb|data| contains the study-specific (i) $S_2$ summary statistics, (ii) mean, standard deviation, and sample size, or (iii) other sets of summary statistics detailed in Appendix A. 
\end{itemize}

For example, the following code applies the $\text{QE}_{\text{median}}$ method to the data set from our example:
\begin{verbatim}
metamedian(data = dat.age, median_method = "qe")
\end{verbatim}
The output is given below.
\begin{verbatim}
Random-Effects Model (k = 51; tau^2 estimator: REML)

tau^2 (estimated amount of total heterogeneity): 33.5585 (SE = 8.3883)
tau (square root of estimated tau^2 value):      5.7930
I^2 (total heterogeneity / total variability):   86.95%
H^2 (total variability / sampling variability):  7.66

Test for Heterogeneity:
Q(df = 50) = 373.6841, p-val < .0001

Model Results:

estimate      se     zval    pval    ci.lb    ci.ub      
 13.2238  0.9121  14.4980  <.0001  11.4361  15.0115  *** 

---
Signif. codes:  0 ‘***’ 0.001 ‘**’ 0.01 ‘*’ 0.05 ‘.’ 0.1 ‘ ’ 1
\end{verbatim}
The output indicates that the pooled estimate of the difference of medians (median age in the nonsurvivor group minus the median age in the survivor group) is 13.22 years [95\% CI: 11.44, 15.01]. 

\subsection{Descriptive analyses} \label{sec: descriptive analyses}

Recall from Section \ref{sec: comparison} that the following factors often strongly affect the performance of the mean-based methods and median-based methods: the proportion of primary studies reporting sample medians, the sets summary statistics reported by the primary studies reporting medians (e.g., $S_1$, $S_2$, or $S_3$), and the skewness of the outcome distribution in the primary studies. The \verb|describe_studies| function prints some descriptive statistics of these factors to help guide data analysts in choosing the most suitable methods to use in their applications. 

Specifically, the \verb|describe_studies| function prints the following information:
\begin{itemize}
    \item The number of primary studies and the number of primary studies reporting medians.
    \item The number of primary studies reporting each set of relevant summary statistics (e.g., $S_1$, $S_2$, and $S_3$). By default, the sets of summary statistics considered are $S_1$, $S_2$ and $S_3$, as well as the sample mean, standard deviation, and sample size. If the argument \verb|method| is set to \verb|"cd"|, the sets of summary statistics considered by the CD median-based method \cite{ozturk2020meta} are used instead (see Appendix A for details). 
    \item The five-number summary (i.e., the minimum value, first quartile, median, third quartile, and maximum value) and mean of the study-specific Bowley skewness values \cite{bowley1901}. Bowley skewness values range from -1 to 1, where positive values indicate that the distribution is right skewed and negative values indicates that the distribution is left skewed. Since the Bowley skewness depends on the sample median and first and third quartiles, the Bowley skewness is only computed for primary studies that report $S_2$ or $S_3$ summary statistics.
\end{itemize}
When the meta-analysis consists of two-group primary studies, the descriptive analyses are performed for each of the two groups. When printing the results, users can specify the labels corresponding to the two groups with the \verb|group_labels| argument.  

For instance, the following code applies the \verb|describe_studies| function to the data set from our example:

\begin{verbatim}
describe_studies(data = dat.age, group_labels = c("Nonsurvivors", "Survivors"))
\end{verbatim}
The output is given below.
\begin{verbatim}
DESCRIPTION OF PRIMARY STUDIES
                                                      Nonsurvivors Survivors
N. studies:                                                     51        51
N. studies reporting the median:                                29        29
  N. studies reporting S1 (min, med, max, n):                    0         0
  N. studies reporting S2 (q1, med, q3, n):                     29        29
  N. studies reporting S3 (min, q1, med, q3, max, n):            0         0
N. studies reporting the mean:                                  22        22
  N. studies reporting the mean, sd, and n:                     22        22
Bowley skewness                                                             
  Minimum:                                                 -0.4000   -0.6000
  First quartile:                                          -0.0818   -0.1304
  Median:                                                   0.0000   -0.0526
  Mean:                                                    -0.0087   -0.0250
  Third quartile:                                           0.0909    0.1458
  Maximum:                                                  0.3846    0.4167
\end{verbatim}

\section{Example} \label{sec: examples}

In this section, we apply the \pkg{metamedian} package to perform a complete analysis of our example meta-analysis comparing the age between survivors and nonsurvivors of COVID-19. 

In these analyses, we do not include one of the primary studies (Qi et al.\ \cite{qi2021clinical}) included in the original meta-analysis, which had a sample size of 5 in the group of nonsurvivors (i.e., we analyze the \verb|dat.age| data set). When including this study, the estimated within-study SE for this study is very large compared to the other primary studies, which can cause instability when estimating between-study heterogeneity. See Appendix B for the results of the analysis when including the study of Qi et al.\ \cite{qi2021clinical}. 

Recall that the descriptive analyses for this data set were performed in Section \ref{sec: descriptive analyses}. Next, we apply several mean-based methods to estimate the pooled difference of mean age between survivors and nonsurvivors. Specifically, we apply all of the applicable mean estimators for $S_2$ summary statistics (i.e., Wan et al.\ \cite{wan2014estimating}, Luo et al.\ \cite{luo2018optimally}, Shi et al.\ \cite{shi2020estimating}, $\text{QE}_{\text{mean}}$ \cite{mcgrath2020estimating}, BC \cite{mcgrath2020estimating}, MLN \cite{cai2021estimating}, and Yang et al.\ \cite{yang2021generalized}). We use the corresponding naïve SE estimator for the Wan et al.\ \cite{wan2014estimating}, Luo et al.\ \cite{luo2018optimally}, and Shi et al.\ \cite{shi2020estimating} mean estimators, the bootstrap SE estimator for the $\text{QE}_{\text{mean}}$ \cite{mcgrath2020estimating}, BC \cite{mcgrath2020estimating}, and MLN \cite{cai2021estimating} mean estimators, and the plug-in SE estimator for the Yang et al.\ \cite{yang2021generalized} mean estimator.  We also apply all of the applicable median-based methods (i.e., the (weighted) MDM method \cite{mcgrath2019one, mcgrath2020meta}, $\text{QE}_{\text{median}}$ \cite{mcgrath2020meta} method) to estimate the pooled difference of median age between survivors and nonsurvivors. For the approaches based on performing an inverse-variance weighted meta-analysis, we assume a random effects model and obtain estimates of the between-study variance and $I^2$ index \cite{higgins2002quantifying, higgins2003measuring} as implemented in the \pkg{metafor} package.

The code for applying the mean-based methods is given below. Due to the computationally intensive nature of bootstrapping and the large number of primary studies, the code took approximately seven minutes to run on a standard laptop computer (8 GB RAM, 1.1 GHz Quad-Core Intel Core i5 processor).
\begin{verbatim}
set.seed(1234)
res_wan <- metamean(dat.age, mean_method = "wan", se_method = "naive", sd_method = "wan")
res_luo <- metamean(dat.age, mean_method = "luo", se_method = "naive", sd_method = "wan")
res_shi <- metamean(dat.age, mean_method = "shi_lognormal", se_method = "naive",
                    sd_method = "shi_lognormal")
res_qe_mean <- metamean(dat.age, mean_method = "qe", se_method = "bootstrap")
res_bc <- metamean(dat.age, mean_method = "bc", se_method = "bootstrap")
res_mln <- metamean(dat.age, mean_method = "mln", se_method = "bootstrap")
res_yang <- metamean(dat.age, mean_method = "yang", se_method = "plugin")
\end{verbatim}
Similarly, the code for applying the median-based methods is given below. 
\begin{verbatim}
res_mm <- metamedian(dat.age, median_method = "mm")
res_wm <- metamedian(dat.age, median_method = "wm")
res_qe_median <- metamedian(dat.age, median_method = "qe")
\end{verbatim}

The results are summarized in Table \ref{tab: application}. Most of the entries in the table can be obtained by printing the objects returned by the \verb|metamean| and \verb|metamedian| functions. The 95\% CIs around $\hat{\tau}^2$ and $\hat{I}^2$ can be obtained by applying the \verb|confint| function from the \pkg{metafor} package (e.g., \verb|confint(res_mln)|). All of the mean-based methods gave similar estimates of the pooled difference of means and between-study heterogeneity, and all of the median-based methods gave similar estimates of the pooled difference of medians. Moreover, the mean-based methods and median-based methods gave similar pooled estimates to each other despite estimating different outcome measures, which presumably occurred because the distribution of age was not highly skewed in the primary studies (e.g., most primary studies had Bowley skewness values smaller than 0.1).

\begin{table}[H]
\caption{Results of the meta-analysis comparing the age between survivors and nonsurvivors of COVID-19. The column titled ``Pooled Estimate [95\% CI]" displays the pooled estimate of the difference of means (mean age in the nonsurvivor group minus the mean age in the survivor group) for the mean-based methods and displays the pooled estimate of the difference of medians for the median-based methods. A value of NA (Not Applicable) is displayed when the method does not provide an estimate of the corresponding parameter. The unit of measurement for age is years. \label{tab: application}}
\begin{center}
\begin{tabular}{llll} \hline
            Method & Pooled Estimate [95\% CI] & $\hat{\tau}^2$ [95\% CI] & $\hat{I}^2$ [95\% CI] \\ \hline
            Mean-Based Methods \\
            \,\,\,\, Wan et al. & 13.34 [11.64, 15.03] & 30.84 [19.19, 50.37] & 89\% [83\%, 93\%] \\
            \,\,\,\, Luo et al. / Wan et al.$^*$ & 13.34 [11.65, 15.04] & 30.69 [19.09, 50.18] & 89\% [83\%, 93\%]\\
            \,\,\,\, Shi et al. & 12.87 [11.25, 14.49] & 27.51 [16.93, 45.66] & 87\% [81\%, 92\%]\\
            \,\,\,\, $\text{QE}_{\text{mean}}$ & 13.23 [11.62, 14.84] & 26.40 [16.02 44.19] & 85\%    [77\%, 90\%] \\
            \,\,\,\, BC & 13.33 [11.72, 14.94] & 25.64 [15.19, 43.53] & 82\% [73\%, 89\%] \\
            \,\,\,\, MLN & 12.86 [11.23, 14.50] & 28.04 [17.35, 46.90] & 87\% [80\%, 92\%] \\
            \,\,\,\, Yang et al. & 13.28 [11.57, 14.99] & 31.28 [19.49, 51.42] & 88\% [82\%, 92\%]\\
            Median-Based Methods \\
            \,\,\,\, MDM & 13.00 [11.72, 16.00] & NA & NA \\
            \,\,\,\, Weighted MDM & 13.00 [8.00, 16.00] & NA & NA  \\ 
            \,\,\,\, $\text{QE}_{\text{median}}$ & 13.22 [11.44, 15.01] & 33.56 [20.70, 55.68] & 87\% [80\% 92\%] \\\hline 
\end{tabular}
\end{center}
\caption*{$^*$We applied the mean estimator of Luo et al. and the standard deviation estimator of Wan et al.}
\end{table}

\begin{figure} [H]
    \centering
    \includegraphics[width=\textwidth]{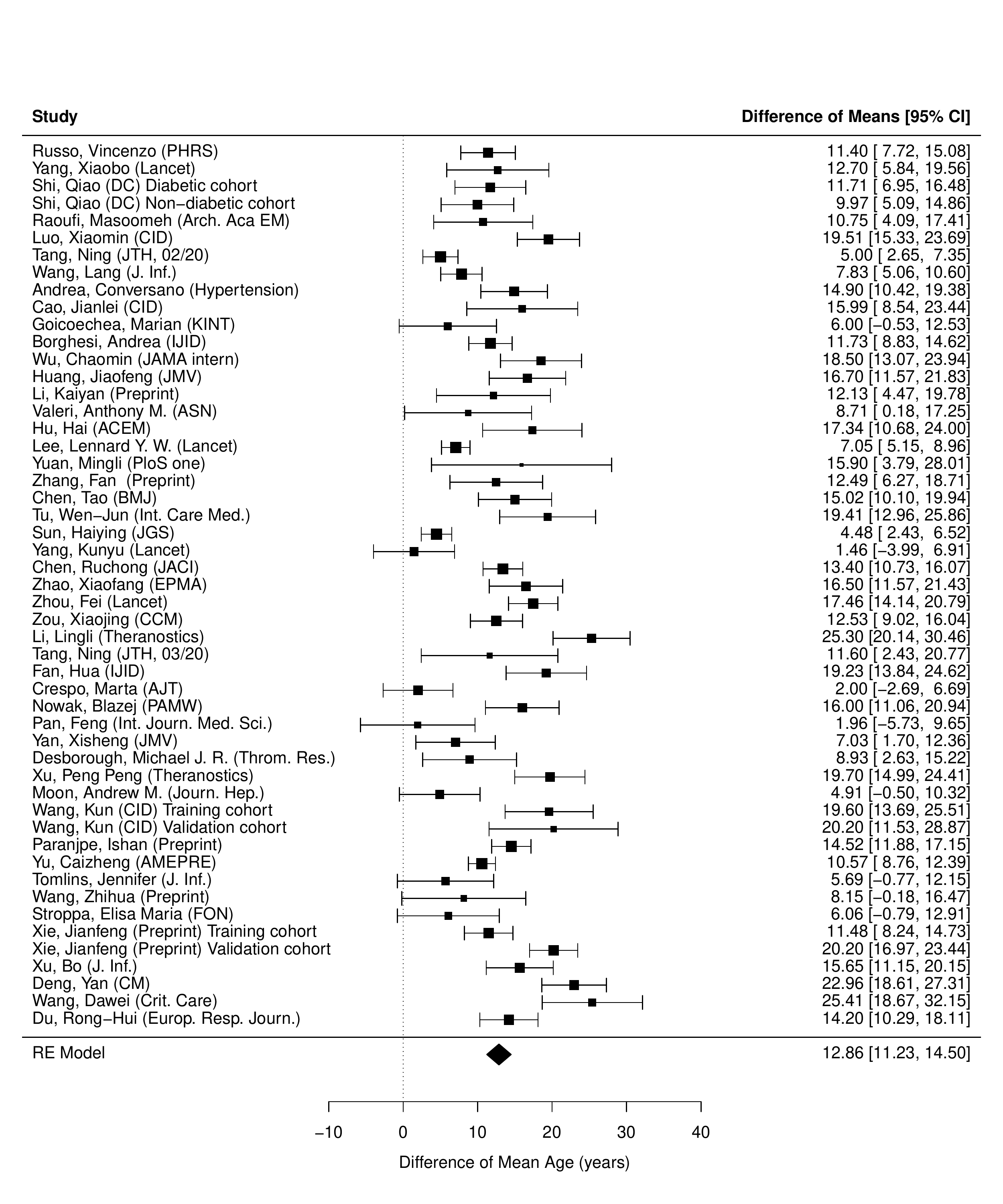}
    \caption{Forest plot showing the difference in mean age between COVID-19 survivors and nonsurvivors. When a primary study reported medians, the MLN method was applied to estimate the difference of means and parametric bootstrap was applied to estimate its standard error \cite{cai2021estimating, mcgrath2023standard}. \label{fig:forest}}
\end{figure}

To generate a forest plot, we can apply the \verb|forest| function from the \pkg{metafor} package. For instance, the code below generates a forest plot corresponding to the analysis with the MLN mean-based method, which is given in Figure \ref{fig:forest}.

\begin{verbatim}
library("metafor")
forest(res_mln, header = c("Study", "Difference of Means [95% CI]"),
       slab = dat.age$author, xlab = "Difference of Mean Age (years)")
\end{verbatim}

\section{Discussion} \label{sec: discussion}

The \pkg{metamedian} R package facilitates performing meta-analyses when primary studies report the sample median of the outcome. The package implements a number of methods that estimate mean-based outcome measures as well as methods that estimate median-based outcome measures. We developed the package in response to the surge of methodological development in this area in recent years and the lack of a comprehensive software tool for such analyses. 

The following subsections discuss some related software, limitations, and future directions.

\subsection{Related software}

In prior work, we developed the \pkg{estmeansd} R package and Shiny web application (\url{https://smcgrath.shinyapps.io/estmeansd/}), which implement the mean and standard deviation estimators proposed by McGrath et al.\ \cite{mcgrath2020estimating} and Cai et al.\ \cite{cai2021estimating} from $S_1$, $S_2$, and $S_3$ summary statistics as well as the corresponding bootstrap SE estimators \cite{mcgrath2023standard}. This package is internally applied in \pkg{metamedian} when applying the relevant mean-based methods. In fact, one motivation for developing \pkg{metamedian} was to facilitate applying \pkg{estmeansd} in meta-analyses and comparing these methods to other mean-based methods and median-based methods in the literature. 

Data analysts should be aware of a few other software options that implement methods to meta-analyze studies reporting medians. The \pkg{meta} R package \cite{balduzzi2019} recently added some mean-based methods (i.e., Wan et al.\ \cite{wan2014estimating}, Luo et al.\ \cite{luo2018optimally}, Shi et al.\ \cite{shi2020optimally}, McGrath et al.\ \cite{mcgrath2020estimating}, and Cai et al.\ \cite{cai2021estimating}, where the latter two are applied via the \pkg{estmeansd} package). The current development version of the \pkg{metafor} R package \cite{metafor} (i.e., version 3.9-21) includes some mean-based methods as well (i.e., Hozo et al.\ \cite{hozo2005estimating}, Walter and Yao \cite{walter2007effect}, Wan et al.\ \cite{wan2014estimating}, Luo et al.\ \cite{luo2018optimally}, and Shi et al.\ \cite{shi2020optimally, shi2020estimating}). Both \pkg{meta} and \pkg{metafor} use the imputed sample means and standard deviations in place of the actual (unreported) sample means and standard deviations in standard meta-analytic approaches (i.e., they use the naïve SE estimator). Last, the \pkg{metaBLUE} R package \cite{metaBLUE} implements the meta-analytic model proposed by Yang et al. \cite{yang2021generalized} to estimate a global mean or global difference of means. 

A key advantage of the \pkg{metamedian} package is that it implements median-based methods, which are currently not available in other software. As previously discussed, prior studies \cite{mcgrath2019one, mcgrath2020meta,mcgrath2023standard} have found that mean-based methods can perform very poorly for highly skewed data and median-based methods may be preferable (and possibly the only option) in such settings. Moreover, by including both mean-based methods and median-based methods in a single software package with the same requirements on formatting the input data set and applying the main functions, we believe that \pkg{metamedian} can greatly facilitate performing sensitivity analyses based on using other methods in the package. A few other advantages of \pkg{metamedian} include the following: (1) \pkg{metamedian} offers within-study bootstrap SE estimators for the mean-based methods, which have been shown to often perform better than the naïve SE estimators. (2) \pkg{metamedian} offers the \verb|describe_studies| function to help data analysts with data cleaning and choosing suitable methods for their application. (3) \pkg{metamedian} includes several real data sets (e.g., \verb|dat.age|, \verb|dat.asat|, \verb|dat.ck|) of meta-analyses consisting of studies reporting medians, which can be useful for methodological development in this area. 

Finally, it should be noted that there are a few software options apart from \pkg{estmeansd} for estimating the sample mean and standard deviation from $S_1$, $S_2$, and $S_3$ summary statistics. Specifically, the \pkg{metaBLUE} R package \cite{metaBLUE} implements the mean and standard deviation estimators of Wan et al. \cite{wan2014estimating}, Luo et al. \cite{luo2018optimally}, and Yang et al. \cite{yang2021generalized}, which is also internally applied in \pkg{metamedian}. The \pkg{ABCMETAapp} Shiny web application \cite{kwon2021abcmetaapp} implements the approximate Bayesian computation (ABC) approach of Kwon and Reis \cite{kwon2015simulation} to estimate the mean and standard deviation.

\subsection{Limitations and future directions}

The \pkg{metamedian} package has a few limitations that should be acknowledged. First, the package is restricted to performing univariate meta-analyses of one-group studies and two-group studies. While some of the methods implemented in the package are applicable in more general settings (e.g., multivariate/multilevel meta-analysis, network meta-analysis, fully Bayesian estimation strategies in meta-analysis), we restricted the package to perform univariate meta-analyses to simplify the user interface. If users are interested in performing meta-analyses that do not fall within the scope of the \pkg{metamedian} package, they can set the \verb|pool_studies| argument (in the \verb|metamean| and \verb|metamedian| functions) to \verb|FALSE| in order to obtain the study-specific outcome measure estimates and their SE estimates which can be used in conjunction with other software options to perform the pooling stage of the analysis. Second, while we strove to make the package as comprehensive as possible, the package does not include some mean and standard deviation estimators developed in this context. For example, we did not include the ABC approach of Kwon and Reis \cite{kwon2015simulation} because it is computationally intensive and its performance can be highly sensitive to a number of settings (e.g., the assumed outcome distribution, prior distributions, acceptance thresholds) that may need to be carefully tailored for each primary study. The mean and standard deviation estimators included in the current version of \pkg{metamedian} are those that we believe users are most likely to apply based on the current literature and that typically do not need to be highly tailored for each primary study.

We aim to expand \pkg{metamedian} in a few of directions in the future. For instance, we would like to include additional outcome measures for the mean-based methods, such as the standardized difference of means and the ratio of means. Similarly, some of the median-based methods can conceivably be extended for other outcome measures (e.g., the ratio of medians). However, we did not include such outcome measures in the current version of \pkg{metamedian} because they first require methodological work. More generally, we anticipate adding new methods to the package as they continue to become developed. We also welcome those working in this area to contribute to the development version of the packages on GitHub.

\section*{Acknowledgements}
We thank Alexandra Zimmer, Alexander Seitel, and Claudia Denkinger for helping collect the data sets used in the examples. This work was supported by the National Science Foundation Graduate Research Fellowship Program under Grant No.\ DGE2140743. Any opinions, findings, and conclusions or recommendations expressed in this material are those of the authors and do not necessarily reflect the views of the National Science Foundation. 

\section*{Conflict of interest}
The authors declare no potential conflict of interest.

\section*{Data availability statement}

The source code of the software and data presented in this paper are publicly available on GitHub (\url{https://github.com/stmcg/metamedian}).

\printbibliography

\end{document}


\maketitle

\appendix 

\section{Additional details on the Confidence Distribution method}

Recall that Ozturk and Balakrishnan \cite{ozturk2020meta} proposed a median-based method for meta-analyzing one-group studies, which they referred to as the Confidence Distribution (CD) method. This method is based on performing an inverse-variance weighted meta-analysis, where one nonparametrically estimates the within-study SEs from various sets for summary statistics. Unlike most of the methods implemented in \pkg{metamedian} -- which consider that primary studies report $S_1$, $S_2$, $S_3$ summary statistics of the outcome or the sample mean, standard deviation, and sample size -- Ozturk and Balakrishnan \cite{ozturk2020meta} consider different sets of summary statistics that may be reported by the primary studies. In this section, we describe these other sets of summary statistics and how to incorporate them when using \pkg{metamedian}. 

To describe the summary statistics considered by Ozturk and Balakrishnan \cite{ozturk2020meta}, we use the following notation. We observe an independent and identically distributed sample of $n$ observations of the outcome of interest, denoted by $\{X_i\}_{i = 1}^n$. Let $X_{(i)}$ denote the $i$th order statistic. Let $\xi$ denote the median of the distribution. Ozturk and Balakrishnan \cite{ozturk2020meta} consider the following sets of summary statistics that may be reported by a primary study: 
\begin{itemize}
    \item[$C_1$:] Two central order statistics, $X_{(r)}$ and $X_{(s)}$ where $1 < r < s < n$, that can form a $100(1-\alpha)\%$ confidence interval (CI) for $\xi$ such that
    \begin{equation} \label{eq: ci}
        P(X_{(r)} \leq \xi \leq X_{(s)}) \geq 1 - \alpha_1 - \alpha_2 = 1 - \alpha
    \end{equation}
    where $P(X_{(r)} \geq \xi) \leq \alpha_1$ and $P(X_{(s)} \leq \xi) \leq \alpha_2$.
    \item[$C_2$:] A $100(1-\alpha)\%$ CI $[L, U] = [X_{(r)}, X_{(s)}]$ for $\xi$ such that (\ref{eq: ci}) holds.
    \item[$C_3$:] The sample median and an estimate of its variance.
    \item[$C_4$:] The sample mean, standard deviation, and $n$.
    \item[$C_5$:] The sample median, first and third quartiles, and $n$.
\end{itemize}

Recall that the \verb|metamedian| function requires an input data set, \verb|data|, containing the summary data from the primary studies. If a primary study reports $C_1$ or $C_2$ summary statistics, users can specify the lower and upper bound of the CI in columns named \verb|med.ci.lb.g1| and \verb|med.ci.ub.g1|, respectively. The $\alpha_1$ and $\alpha_2$ values can be specified in columns named \verb|alpha.1.g1| and \verb|alpha.2.g1|, respectively. If a primary study reports $C_3$ summary statistics, users can specify the estimated variance of the median in a column named \verb|med.var.g1| (and can specify the median in a column named \verb|med.g1|). Recall that the main text discusses how to specify the $C_4$ and $C_5$ summary statistics in the input data set (i.e., in columns named \verb|mean.g1|, \verb|sd.g1|, \verb|n.g1|, \verb|med.g1|, \verb|q1.g1|, and \verb|q3.g1|). 

\section{Additional results from the main example}

\setcounter{table}{0}
\renewcommand{\thetable}{B\arabic{table}}

Table \ref{tab: application with qui} summarizes the results of the analysis of our main example when including the study of Qi et al.\ \cite{qi2021clinical}. The code for performing the analysis is the same as that presented in Section 4 of the main text, replacing the data set \verb|dat.age| with \verb|dat.age_raw|.

\begin{table}[H]
\caption{Results of the meta-analysis comparing the age between survivors and nonsurvivors of COVID-19 when including the study of Qi et al.\ \cite{qi2021clinical}. The column titled ``Pooled Estimate [95\% CI]" displays the pooled estimate of the difference of means (mean age in the nonsurvivor group minus the mean age in the survivor group) for the mean-based methods and displays the pooled estimate of the difference of medians for the median-based methods. A value of NA (Not Applicable) is displayed when the method does not provide an estimate of the corresponding parameter. The unit of measurement for age is years.\label{tab: application with qui}}
\begin{center}
\begin{tabular}{llll} \hline
            Method & Pooled Estimate [95\% CI] & $\hat{\tau}^2$ [95\% CI] & $\hat{I}^2$ [95\% CI] \\ \hline
            Mean-Based Methods \\
            \,\,\,\, Wan et al. & 13.28 [11.58, 14.97] & 30.86 [19.20, 50.77] & 89\% [83\%, 93\%] \\
            \,\,\,\, Luo et al. / Wan et al.$^*$ & 13.28 [11.59, 14.97] & 30.71 [19.12, 50.66] & 89\% [83\%, 93\%]\\
            \,\,\,\, Shi et al. & 12.81 [11.19, 14.43] & 27.54 [16.98, 46.17] & 87\% [81\%, 92\%]\\
            \,\,\,\, $\text{QE}_{\text{mean}}$ & 13.15 [11.55, 14.76] & 26.56 [16.26 45.30] & 84\%    [77\%, 90\%] \\
            \,\,\,\, BC & 13.33 [11.72, 14.93] & 25.30 [14.53, 41.57] & 82\% [72\%, 88\%] \\
            \,\,\,\, MLN & 12.77 [11.16, 14.39] & 27.44 [16.93, 46.18] & 86\% [80\%, 91\%] \\
            \,\,\,\, Yang et al. & 13.23 [11.52, 14.93] & 31.26 [19.40, 51.43] & 88\% [82\%, 92\%]\\
            Median-Based Methods \\
            \,\,\,\, MDM & 13.00 [11.51, 16.00] & NA & NA \\
            \,\,\,\, Weighted MDM & 13.00 [8.00, 16.00] & NA & NA  \\ 
            \,\,\,\, $\text{QE}_{\text{median}}$ & 13.15 [11.37, 14.94] & 33.59 [20.74, 56.16] & 87\% [80\% 92\%] \\\hline 
\end{tabular}
\end{center}
\caption*{$^*$We applied the mean estimator of Luo et al. and the standard deviation estimator of Wan et al.}
\end{table}

\section{Additional examples}

In this section, we analyze two additional example data sets from the meta-analysis of Katzenschlager et al.\ \cite{katzenschlager2021can}. Specifically, we meta-analyze aspartate transaminase (ASAT) and creatine kinase (CK) levels in the group of COVID-19 nonsurvivors. 

We chose these additional examples for two main reasons: 1) Unlike the outcome of age in the main example, the distribution of ASAT levels is moderately skewed and the distribution of CK levels is highly skewed in most of the primary studies. When meta-analyzing skewed outcomes, the different methods in the \pkg{metamedian} package can lead to considerably different results. This underscores the importance of \pkg{metamedian} allowing data analysts to choose from various methods and to easily perform sensitivity analyses with different methods. 2) These examples are meta-analyses of single group data, unlike the main example which meta-analyzes two-group data. This allows us to illustrate using \pkg{metamedian} to meta-analyze other outcome measures (i.e., the mean and median of the outcome) and to apply the CD method \cite{ozturk2020meta}.

\subsection{Aspartate transaminase levels in COVID-19 nonsurvivors}

\setcounter{table}{0}
\renewcommand{\thetable}{C\arabic{table}}

\setcounter{figure}{0}
\renewcommand{\thefigure}{C\arabic{figure}}

The data sets \verb|dat.asat| and \verb|dat.asat_raw| in \pkg{metamedian} contain the summary data of ASAT levels in COVID-19 survivors and nonsurvivors, where \verb|dat.asat_raw| includes the primary study of Qi et al.\ \cite{qi2021clinical} and \verb|dat.asat| does not. For simplicity, we use \verb|dat.asat| in this subsection.

The first five rows of \verb|dat.asat| (excluding the \verb|author| column) are displayed below. The group 1 values correspond to nonsurvivor group, and the group 2 values correspond to the survivor group. The unit of measurement for the ASAT levels is U/L.
\begin{verbatim}
   n.g1 q1.g1 med.g1 q3.g1 mean.g1  sd.g1 n.g2 q1.g2 med.g2  q3.g2 mean.g2  sd.g2
1    29    NA     NA    NA  104.56 117.76  351    NA     NA     NA   64.38 234.43
2    65 30.00   43.0  68.0      NA     NA  274 22.00   29.0  43.00      NA     NA
3    12 27.25   38.0  70.5      NA     NA    9 36.50  112.0 293.25      NA     NA
4    13 28.75   38.0  64.5      NA     NA   40 20.00   25.0  35.00      NA     NA
5    44 30.00   37.0  52.0      NA     NA   40 32.25   38.5  57.25      NA     NA
\end{verbatim}

Since \verb|dat.asat| includes columns for the survivor group, we need to subset \verb|dat.asat| in order for the \verb|metamean| and \verb|metamedian| functions to meta-analyze the mean and median ASAT levels, respectively, in the nonsurvivor group. This can be performed as follows:
\begin{verbatim}
my_data <- dat.asat[, c("author", "q1.g1", "med.g1", "q3.g1", "mean.g1", "sd.g1", "n.g1")]
\end{verbatim}

Next, we perform the descriptive analyses by applying the command \verb|describe_studies(my_data)|, which produces the following output:
\begin{verbatim}
DESCRIPTION OF PRIMARY STUDIES
                                                             
N. studies:                                                26
N. studies reporting the median:                           23
  N. studies reporting S1 (min, med, max, n):               0
  N. studies reporting S2 (q1, med, q3, n):                23
  N. studies reporting S3 (min, q1, med, q3, max, n):       0
N. studies reporting the mean:                              3
  N. studies reporting the mean, sd, and n:                 3
Bowley skewness                                              
  Minimum:                                            -0.1917
  First quartile:                                      0.1901
  Median:                                              0.2903
  Mean:                                                0.2686
  Third quartile:                                      0.3818
  Maximum:                                             0.5556
\end{verbatim}
Observe that most primary studies report $S_2$ summary statistics and have Bowley skewness values greater than 0.2, indicating a moderate degree of right skewness in the distribution of ASAT levels.

Next, we apply the same mean-based methods and median-based methods as described in Section 4 of the main text. In addition, since this example is a meta-analysis of single group data, we also apply the CD method \cite{ozturk2020meta} in a random effects meta-analysis. The code for performing these analyses is given below. The code took approximately three minutes to run on the same computer as described in Section 4 of the main text.

\begin{verbatim}
## Applying mean-based methods
set.seed(1234)
res_wan <- metamean(my_data, mean_method = "wan", se_method = "naive", sd_method = "wan")
res_luo <- metamean(my_data, mean_method = "luo", se_method = "naive", sd_method = "wan")
res_shi <- metamean(my_data, mean_method = "shi_lognormal", se_method = "naive", 
                    sd_method = "shi_lognormal")
res_qe_mean <- metamean(my_data, mean_method = "qe", se_method = "bootstrap")
res_bc <- metamean(my_data, mean_method = "bc", se_method = "bootstrap")
res_mln <- metamean(my_data, mean_method = "mln", se_method = "bootstrap")
res_yang <- metamean(my_data, mean_method = "yang", se_method = "plugin")

## Applying median-based methods
res_mm <- metamedian(my_data, median_method = "mm")
res_wm <- metamedian(my_data, median_method = "wm")
res_qe_median <- metamedian(my_data, median_method = "qe")
res_cd <- metamedian(my_data, median_method = "cd")
\end{verbatim}

Table \ref{tab: application asat} summarizes the results. The mean-based methods that assume that the outcome is normally distributed (i.e., Wan et al.\ \cite{wan2014estimating}, Luo et al.\ \cite{luo2018optimally}, and Yang et al.\ \cite{yang2021generalized}) gave pooled mean estimates of approximately 46 U/L, whereas the mean-based methods that do not assume normality (i.e., Shi et al.\ \cite{shi2020estimating}, $\text{QE}_{\text{mean}}$ \cite{mcgrath2020estimating}, BC \cite{mcgrath2020estimating}, and MLN \cite{cai2021estimating}) gave pooled mean estimates of approximately 50 U/L. All of the median-based methods gave pooled median estimates of approximately 42 U/L. The observed differences in the pooled estimates between the various methods may be attributed to ASAT levels being right skewed in many of the primary studies. 

The mean-based methods assuming normality had smaller estimates of $\tau^2$ compared to those that do not assume normality. The trends in the $I^2$ estimates were less clear, which may be attributed to $I^2$ strongly depending on both the distributional assumptions of the mean-based method as well as the choice of the within-study SE estimator. Moreover, all of the mean-based methods had larger estimates of between-study heterogeneity compared to the median-based methods.

\begin{table}[H]
\caption{Results of the meta-analysis of aspartate transaminase (ASAT) levels in COVID-19 nonsurvivors. The column titled ``Pooled Estimate [95\% CI]" displays the pooled estimate of mean ASAT level (in U/L) for the mean-based methods and displays the pooled estimate of the median ASAT level (in U/L) for the median-based methods. A value of NA (Not Applicable) is displayed when the method does not provide an estimate of the corresponding parameter. \label{tab: application asat}}
\begin{center}
\begin{tabular}{llll} \hline
            Method & Pooled Estimate [95\% CI] & $\hat{\tau}^2$ [95\% CI] & $\hat{I}^2$ [95\% CI] \\ \hline
            Mean-Based Methods \\
            \,\,\,\, Wan et al. & 45.93 [42.35, 49.50] & 57.20 [37.68, 286.85] & 82\% [76\%, 96\%] \\
            \,\,\,\, Luo et al. / Wan et al.$^*$ & 46.13 [42.53, 49.74] & 58.36 [38.20, 287.27] & 83\% [76\%, 96\%]\\
            \,\,\,\, Shi et al. & 51.37 [46.57, 56.17] & 112.94 [75.70, 484.90] & 88\% [83\%, 97\%]\\
            \,\,\,\, $\text{QE}_{\text{mean}}$ & 49.45 [45.14, 53.76] & 73.93 [42.08 365.46] & 76\%    [64\%, 94\%] \\
            \,\,\,\, BC & 49.57 [45.27, 53.87] & 69.90 [35.46, 332.15] & 73\% [58\%, 93\%] \\
            \,\,\,\, MLN & 49.86 [45.39, 54.32] & 92.16 [60.54, 429.83] & 83\% [77\%, 96\%] \\
            \,\,\,\, Yang et al. & 45.76 [42.21, 49.31] & 55.74 [35.60, 275.11] & 81\% [73\%, 95\%]\\
            Median-Based Methods \\
            \,\,\,\, MDM & 41.50 [38.00, 43.89] & NA & NA \\
            \,\,\,\, Weighted MDM & 43.40 [42.00, 47.00] & NA & NA  \\ 
            \,\,\,\, $\text{QE}_{\text{median}}$ & 42.58 [39.58, 45.58] & 30.50 [20.36, 264.74] & 63\% [54\% 94\%] \\
            \,\,\,\, CD & 42.53 [39.41, 45.65] & 34.26 [NA, NA] & NA \\\hline 
\end{tabular}
\end{center}
\caption*{$^*$We applied the mean estimator of Luo et al. and the standard deviation estimator of Wan et al.}
\end{table}

Last, we generate a forest plot corresponding to the analysis with the MLN mean-based method. The code for generating the forest plot is given below, and the forest plot is given in Figure \ref{fig:forest asat}.

\begin{verbatim}
forest(res_mln, header = c("Study", "Mean [95% CI]"),
       slab = my_data$author, xlab = "Mean ASAT (U/L)")
\end{verbatim}

\begin{figure} [H]
    \centering
    \includegraphics[width=\textwidth]{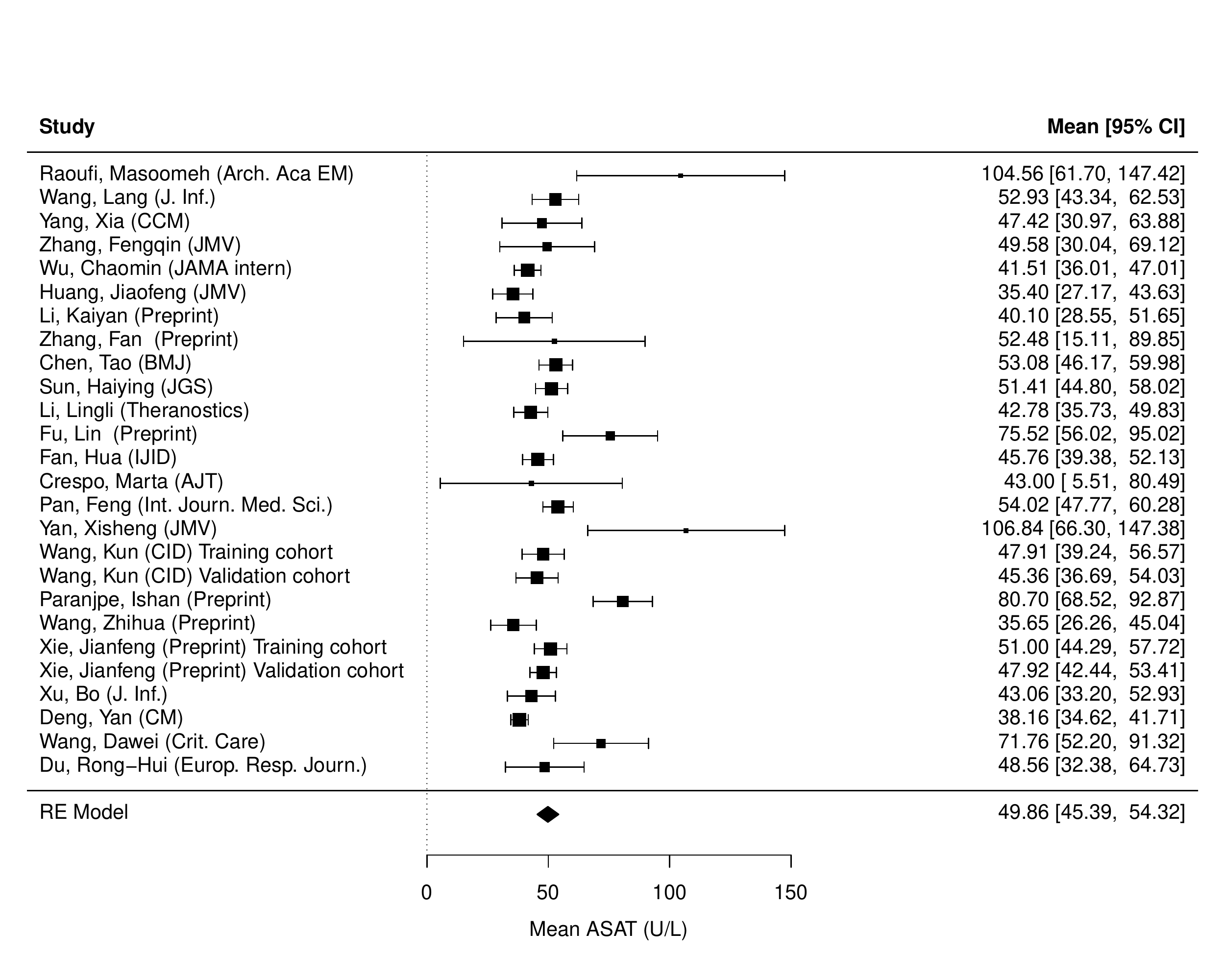}
    \caption{Forest plot showing the mean ASAT level in COVID-19 nonsurvivors. When a primary study reported the median, the MLN method was applied to estimate the mean and parametric bootstrap was applied to estimate its standard error \cite{cai2021estimating, mcgrath2023standard}. \label{fig:forest asat}}
\end{figure}

\subsection{Creatine kinase levels in COVID-19 nonsurvivors}

The data sets \verb|dat.ck| and \verb|dat.ck_raw| in \pkg{metamedian} contain the summary data of CK levels in COVID-19 survivors and nonsurvivors. Similar to the previous example, \verb|dat.ck_raw| includes the primary study of Qi et al.\ \cite{qi2021clinical}. Once again, we use the data set excluding the primary study of Qi et al.\ \cite{qi2021clinical} in this subsection.

The first five rows of \verb|dat.ck| (excluding the \verb|author| column) are below. As in the previous examples, the group 1 values correspond to the nonsurvivor group, and the group 2 values correspond to the survivor group. The unit of measurement for the CK levels is U/L.
\begin{verbatim}
   n.g1  q1.g1 med.g1 q3.g1 mean.g1   sd.g1 n.g2 q1.g2 med.g2   q3.g2 mean.g2  sd.g2
1    29     NA     NA    NA 1033.45 1754.19  351    NA     NA      NA  216.38 474.51
2    65  50.00   84.0 222.0      NA      NA  274  40.0   60.0   97.00      NA     NA
3    12  38.75   96.0 415.5      NA      NA    9  63.0  100.5 2322.75      NA     NA
4    48 106.70  251.0 392.0      NA      NA  185  59.0   93.0  182.20      NA     NA
5    11     NA     NA    NA   57.00   36.80   25    NA     NA      NA   85.00  69.00
\end{verbatim}

As in the previous example, we need to subset \verb|dat.ck| in order to meta-analyze the mean and median CK level in the nonsurvivor group. The code for performing this is the same as in the previous example when replacing \verb|dat.asat| with \verb|dat.ck|. Moreover, since the code for performing the descriptive analyses and the meta-analyses is identical to that in the previous example after subsetting the data, we do not display the code throughout this example for parsimony. 

The output of the descriptive analyses is given below:
\begin{verbatim}
DESCRIPTION OF PRIMARY STUDIES
                                                             
N. studies:                                                17
N. studies reporting the median:                           15
  N. studies reporting S1 (min, med, max, n):               0
  N. studies reporting S2 (q1, med, q3, n):                15
  N. studies reporting S3 (min, q1, med, q3, max, n):       0
N. studies reporting the mean:                              2
  N. studies reporting the mean, sd, and n:                 2
Bowley skewness                                              
  Minimum:                                            -0.0116
  First quartile:                                      0.2934
  Median:                                              0.3840
  Mean:                                                0.4381
  Third quartile:                                      0.6076
  Maximum:                                             0.8416
\end{verbatim}
Observe that most primary studies report $S_2$ summary statistics and that the skewness of the outcome in the primary studies is somewhat larger than that of the previous example. Most studies had Bowley skewness values greater than 0.3, and one study (Zhang et al.\ \cite{zhang2020obesity}) had a Bowley skewness value as large as 0.84.

Next, we apply the same methods as described in the previous example to meta-analyze the mean and median CK level. The code for performing the analyses took approximately two minutes to run on the same computer as described in Section 4 of the main text. 

Table \ref{tab: application ck} summarizes the results. The mean-based methods that assume that the outcome is normally distributed gave pooled estimates ranging from 181 U/L to 187 U/L, whereas the mean-based methods that do not assume normality gave pooled mean estimates ranging from 221 U/L to 257 U/L. All the median-based methods gave similar pooled median estimates of approximately 140 U/L. As in the previous example, the discrepancies between the pooled estimates of these methods are presumably due to the skewness of the distribution of CK levels in the primary studies. 

The estimates of between-study heterogeneity were very large for all of the methods. The trends in the estimates of between-study heterogeneity for the various methods were less clear in this example compared to the previous one. 

The results for the $\text{QE}_{\text{mean}}$ and BC mean-based methods should be interpreted with some caution. Both of these methods had very large within-study SE estimates for the primary study of Yang et al.\ \cite{yang2020extracorporeal}, and the $\text{QE}_{\text{mean}}$ method additionally had a very large within-study SE estimate for the primary study of Zhang et al.\ \cite{zhang2020obesity}. These two primary studies had very small sample sizes ($n = 8$ and $n = 12$, respectively), and the bootstrap SE estimators have been found to perform poorly for such small sample sizes \cite{mcgrath2023standard}. The presence of very large within-study SE estimates can cause instability when estimating between-study heterogeneity. 

\begin{table}[H]
\caption{Results of the meta-analysis of creatine kinase (CK) levels in COVID-19 nonsurvivors. The column titled ``Pooled Estimate [95\% CI]" displays the pooled estimate of mean CK level (in U/L) for the mean-based methods and displays the pooled estimate of the median CK level (in U/L) for the median-based methods. A value of NA (Not Applicable) is displayed when the method does not provide an estimate of the corresponding parameter. \label{tab: application ck}}
\begin{center}
\begin{tabular}{llll} \hline
            Method & Pooled Estimate [95\% CI] & $\hat{\tau}^2$ [95\% CI] & $\hat{I}^2$ [95\% CI] \\ \hline
            Mean-Based Methods \\
            \,\,\,\, Wan et al. & 184.77 [134.21, 235.34] & 9272.31 [5962.93, 66248.33] & 95\% [92\%, 99\%] \\
            \,\,\,\, Luo et al. / Wan et al.$^*$ & 187.16 [136.38, 237.95] & 9365.94 [6004.80, 66146.39] & 95\% [92\%, 99\%]\\
            \,\,\,\, Shi et al. & 244.54 [180.69, 308.38] & 12499.12 [9871.40, 138972.32] & 92\% [90\%, 99\%]\\
            \,\,\,\, $\text{QE}_{\text{mean}}$ & 257.12 [175.24, 339.00] & 13536.63 [3262.90 70239.81] & 88\%    [64\%, 97\%] \\
            \,\,\,\, BC & 234.82 [169.24, 300.41] & 9108.76 [2642.80, 57973.02] & 84\% [60\%, 97\%] \\
            \,\,\,\, MLN & 220.69 [160.29, 281.09] & 10015.95 [6184.21, 120680.56] & 90\% [85\%, 99\%] \\
            \,\,\,\, Yang et al. & 180.91 [131.54, 230.28] & 8760.71 [5733.60, 66314.43] & 94\% [92\%, 99\%]\\
            Median-Based Methods \\
            \,\,\,\, MDM & 137.00 [97.68, 188.41] & NA & NA \\
            \,\,\,\, Weighted MDM & 142.00 [137.00, 189.00] & NA & NA  \\ 
            \,\,\,\, $\text{QE}_{\text{median}}$ & 145.42 [102.61, 188.23] & 6187.63 [4436.58, 66290.36] & 91\% [88\% 99\%] \\
            \,\,\,\, CD & 140.57 [96.20, 184.93] & 4121.38 [NA, NA] & NA \\\hline 
\end{tabular}
\end{center}
\caption*{$^*$We applied the mean estimator of Luo et al. and the standard deviation estimator of Wan et al.}
\end{table}

\begin{figure} [H]
    \centering
    \includegraphics[width=\textwidth]{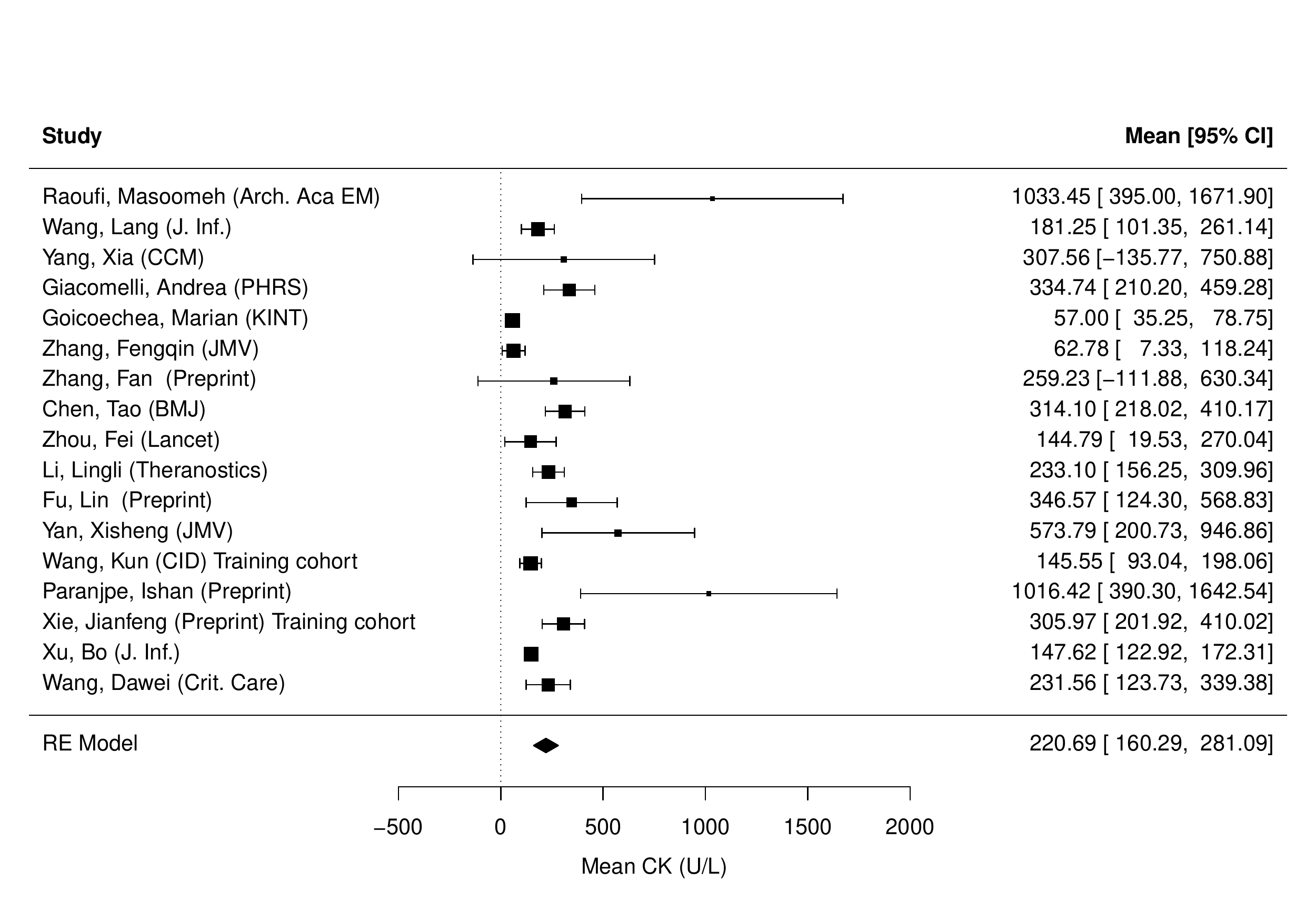}
    \caption{Forest plot showing the mean CK level in COVID-19 nonsurvivors. When a primary study reported the median, the MLN method was applied to estimate the mean and parametric bootstrap was applied to estimate its standard error \cite{cai2021estimating, mcgrath2023standard}. \label{fig:forest ck}}
\end{figure}

For consistency with the previous examples, we illustrate a forest plot of the analysis based on the MLN mean-based method in Figure \ref{fig:forest ck}.

\printbibliography